\documentclass[twocolumn,showpacs,aps]{revtex4}

\usepackage{graphicx}

\begin{document}

\title{Anharmonic vs. relaxational sound damping in glasses:\\	
I. Brillouin scattering from densified silica}

\author{Evelyne Rat$^1$}
\author{Marie Foret$^2$}
 \email[Corresponding author: ]{Marie.Foret@lcvn.univ-montp2.fr}
\author{Gladys Massiera$^1$}
 \altaffiliation[Present address: ]{Laboratoire de Spectrom\'etrie Physique, Universit\'e Joseph Fourier, F-38402 Saint Martin d'H\`eres, France.}
\author{R\'emy Vialla$^2$}
\author{Masatoshi Arai$^3$}
\author{Ren\'e Vacher$^2$}
\author{Eric Courtens$^2$}

\affiliation{$^1$Laboratoire des Verres, UMR 5587 CNRS, Universit\'e Montpellier II, F-34095 Montpellier Cedex 5, France\\
$^2$Laboratoire des Collo\"{\i}des, Verres et Nanomat\'eriaux, UMR 5587 CNRS, Universit\'e Montpellier II, F-34095 Montpellier Cedex 5, France\\
$^3$High Intensity Proton Accelerator Project Division, High Energy Accelerator Research Organization, 1-1 Oho, Tsukuba 305-0801, Japan}

\date{\today}

\begin{abstract}
This series discusses the origin of sound damping and dispersion
in glasses.
In particular, we address the relative importance of anharmonicity versus
thermally activated relaxation.
In this first article, Brillouin-scattering measurements of permanently
densified silica glass are presented.
It is found that in this case the results are compatible with a model in which
damping and dispersion are only produced by the anharmonic coupling
of the sound waves with thermally excited modes.
The thermal relaxation time and the unrelaxed velocity are estimated.
\end{abstract}


\maketitle

\section{Introduction}

The understanding of the physical processes leading to sound attenuation
and dispersion in dielectric glasses is still far from perfect.
At least four distinct causes have been discussed: 
two-level systems \cite{And72,Phi72,Jae72,Phi81}, relaxation
\cite{And55,Hun76,Par94},
Rayleigh scattering \cite{Ray96,Kle51,Ell92}, and anharmonicity.
Anharmonicity produces a direct
interaction of sound with thermal vibrations.
One knows that for dielectric {\em crystals}
it is a dominant contribution to the damping of acoustic waves
\cite{Mar71}.
This has been called either ``lattice viscosity'' \cite{Vau67} or
``phonon viscosity'' \cite{Mar69}.
In Akhiezer's treatment \cite{Akh39,Pom41}, the sound wave modulates the
frequencies of the thermally excited modes and thereby their
effective temperatures.
The return to thermal equilibrium within this vibrational bath, characterized by the mean
thermal lifetime of the dominant vibrations $\tau_{\rm th}$, leads then to energy dissipation
and thereby to sound damping.
This process can be quite local, occuring between spatially
overlapping but not necessarily propagating modes \cite{Boe60}.
The time $\tau_{\rm th}$ just describes the return to equilibrium after
a perturbation of these vibrations by an acoustic wave.
Described in these terms, the Akhiezer mechanism also ought to be active
in glasses, even though the thermal modes might then be local or diffusive ones,
leading to what we shall call ``network viscosity''.
One of us had already suggested long ago that at the high frequencies
of Brillouin scattering experiments, this Akhiezer contribution should not
be neglected in vitreous silica \cite{Vac81}. A simulation based on this mechanism
was performed for the case of amorphous silicon \cite{Fab99}. It showed that the
anharmonicity of thermal vibrations can induce in glasses sound attenuation larger
than in the corresponding crystal owing to the enhancement of the Gr\"uneisen parameters
by internal strains. Other detailed calculations of the anharmonic damping of sound
in glasses do not seem available in the literature,
except for extensive predictions of relaxation times based on the fracton model
\cite{Ale86}.

At temperatures above the quantum regime, owing to the considerable
disorder of glasses, damping can also result from unstable
structural features relaxing in the strain field of the sound wave
\cite{And55,Hun76}.
These ``defects'' produce what is called ``thermally activated relaxation''.
The damping depends then
on the defect density, on the strength of their coupling to strain, and
on the distribution of their relaxation times $\tau$.
These times are often much longer than typical values
of $\tau_{\rm th}$.
Hence, this type of damping appears dominant in many glasses at sonic and
ultrasonic frequencies.
It produces peaks in the
temperature dependence of the sound attenuation which have
been well described by phenomenology \cite{Hun76}, although
the exact nature of the defects remains most often unknown.
A puzzling aspect is the relative ``universality'' of these relaxation
phenomena, as reviewed in \cite{Poh02}.
Another non-trivial aspect is that within the soft-potential model the
relaxational features at higher temperatures and the properties associated
with two-level systems, at He temperatures and below, can be quite succesfully
described using a single picture \cite{Par94}.

Another possible damping mechanism is scattering
of the sound waves by static density or elastic constant fluctuations.
This effect which grows with the fourth power of the frequency
was described long ago by Lord Rayleigh \cite{Ray96}.
Although this mechanism has been invoked to explain the plateau in the thermal
conductivity of glasses, or the ``end'' of acoustic
branches \cite{Car61,Gra86,Ell92,Rat99},
it has been known for quite some time that it is difficult to construct a model
for glasses that produces strong Rayleigh scattering \cite{Zai75}.
It has been recently reemphasized that the Rayleigh scattering of sound
is generally too weak for it to play any appreciable role at
frequencies below at least 1 THz \cite{Par01,Gur03}.
Thus, this will not be of importance to the present series of
papers as our considerations will be restricted to frequencies below
$\sim 100$ GHz.
Similarly, another very strong damping mechanism is the resonant
interaction --or hybridization-- of the sound with other low frequency
modes, which can lead to ``boson peaks'' \cite{Kar85,Kli01}.
In the case of densified silica, this effect has been observed to start
around 1 THz and to grow then with a high power of the sound frequency
\cite{Ruf03,Cou03}.
Again, this will not be of relevance here.
Our interest will in fact mainly be in the relative strength of the
anharmonic and relaxation sources of damping.
At very low temperatures, below liquid He, the quantum aspect of
two-level systems becomes crucial {\cite{Jae72,Vac80} but anharmonic
damping tends to zero
owing to the weak population of the thermal bath.
Very low temperatures will not
be discussed here.

Compared to the very abundant ultrasonic results there exist
relatively few damping data at frequencies
between 1 GHz and 1 THz.
The ``hypersonic'' regime observed with
optical Brillouin scattering falls in the middle of this range.
The longitudinal acoustic (LA) waves are active in the backscattering geometry.
To fix ideas, in the case of silica and using the
green argon-laser line, the observed
hypersonic frequency is then $\nu_{\rm B} \equiv \Omega / 2 \pi \simeq 34$ GHz.
Explanations that are only based on thermally activated relaxations generally
meet difficulties in trying to account simultaneously for hypersonic
and ultrasonic attenuation results, {\em e.g.} in \cite{Her98}.
However, it was recently claimed that for silica, ``within a factor of two'',
the results over the entire frequency range are well accounted for by
relaxation only \cite{Wie00}.
A main purpose of the present series is to clarify this issue.
This becomes all the more important that data are now also
obtained at frequencies above $\sim 250$ GHz, using
x-ray Brillouin scattering \cite{Ruo01},\cite{Ruf03}.
In that context, one should beware of sweeping interpolations
of the damping behavior over great many decades in frequency \cite{Ruo01}.
As explained in the second paper of this series (II),
the damping contribution of anharmonicity in
the hypersonic regime increases faster in $\Omega$ than that of
relaxations.
At sufficiently high frequencies and temperatures $T$, this can lead,
over certain frequency and temperature ranges, to
a dominance of anharmonicity.
Indeed, the thermal phonon contribution
increases with the thermal population and thus with some power of $T$,
whereas the effect of the relaxing defects usually saturates.

In order to estimate the anharmonic contribution to sound dispersion
and damping, it is of interest to consider several varieties of silica
glasses besides usual vitreous silica, $v$-SiO$_2$.
For example, it is already known that irradiation of $v$-SiO$_2$ by
fast neutrons increases slightly the glass density $\rho$, from 2.20 to
$\sim 2.26$ g/cm$^3$ \cite{Luk55}, but thereby
increases appreciably the hypersonic velocity while it reduces
substantially the hypersonic attenuation \cite{Bon94}.
Similarly, it has already been shown that in permanently densified silica glass
both the relaxation at sonic frequencies \cite{Wei96} and the hypersonic
linewidth \cite{Doe94} are very strongly reduced.
The latter experiments were performed on a sample of mean density
$\rho \simeq 2.45$ g/cm$^3$ which unfortunately was quite inhomogeneous.
It is thus of considerable interest to obtain detailed $T$-dependent
Brillouin data on a much better sample of well densified silica,
$d$-SiO$_2$, with $\rho \simeq 2.60$ g/cm$^3$.
The results of such an experiment are presented and discussed
in this first paper.
They confirm that the hypersonic damping in $d$-SiO$_2$ is indeed
entirely dominated by anharmonicity.
This is of particular interest as it reveals for the first time
what the shape and strength
of a pure anharmonic contribution can be in a glass.

Two quantities will be of main interest in these papers, the
damping of sound and its velocity dispersion.
Damping can be expressed in several ways.
It is manifested in a Brillouin scattering measurement by a linewidth
$\Gamma$, where $\Gamma /2 \pi$ is defined as the deconvoluted
frequency half-width at half-maximum of
a damped harmonic oscillator response adjusted to the Brillouin peak.
On the other hand, in ultrasonics the damping is generally measured as an
{\em energy} mean-free path $\ell$, or as its inverse $\ell ^{-1} = \alpha =
2 \Gamma / v$, where $v$ is the sound velocity and $\Gamma$ is in rad/s.
Traditionally, the attenuation constant $\alpha$ is often expressed
in dB/cm, in which case its numerical value must be multiplied
by $10 \: ln \: 10$ to convert it to m$^{-1}$.
To compare these various results over a broad range of frequencies,
we prefer to use the internal friction $Q^{-1}$ which is the inverse of the
quality factor $Q= \Omega / 2 \Gamma$.
Thus, $Q^{-1}=\ell ^{-1} v / \Omega$.
This is particularly convenient for graphical presentation as near
a relaxation peak $Q^{-1}$ depends little on
$\Omega$, this over many decades of $\Omega$.
The dispersion in $v(\Omega , T)$ is usually quite weak, and thus
it is indicated to present a differential quantity, $\delta v / v$.
In defining this quantity a suitable reference value $v_0$ must be selected,
as $\delta v / v$ really means $(v(\Omega , T)-v_0)/v_0$.
A rather common choice is to take for $v_0$ the value of $v(\Omega ,T)$ at the
lowest measured temperature, usually near liquid-He temperatures.
One should be aware that an arbitrary choice of $v_0$
modifies $\delta v / v$ practically by an additive constant, since $\delta v \ll v$.
Further, if  $Q^{-1}$ is small compared to 1, which is always
true here, one easily shows that $Q^{-1}$ and $-2 \delta v / v$ are
Kramers-Kronig transforms of each other,
$$-2 \delta v(\Omega ,T) / v \; = \;
\frac{1}{\pi} \; P\int_{-\infty}^{+\infty}\frac
{Q^{-1}(x,T)}{x - \Omega} \; dx \; , \eqno{(1)}$$
with the related relation for the inverse transform.
As explained in \cite{LL}, this strictly requires that
$\delta v = v - v_\infty$, where $v_\infty = v(\Omega \rightarrow \infty,T)$ is the high-frequency limit of $v$,
which can be called the unrelaxed velocity.
It is conceivable that ``static'' structural changes might occur in function of $T$,
and that these could produce a temperature dependent $v_\infty (T)$.

As discussed in \cite{Jae76}, if $Q^{-1}$ is given by a single Debye relaxation,
$$ Q^{-1}(\Omega ,T) = A \Omega \tau / (1+ \Omega ^2 \tau ^2 ) \; , \eqno{(2a)}$$
where $A$ and $\tau$ are independent of frequency but might depend on $T$,
it follows from (1) that
$$ -2 \delta v (\Omega ,T) / v = A / (1+ \Omega ^2 \tau ^2 ) \; . \eqno{(2b)}$$
Finally, concerning the various independent contributions to $\Gamma$, and
thus to $Q^{-1}$, we shall adopt the equivalent of ``Matthiessen's rule''
\cite{Kit76} that these are simply additive.
Thus, if $Q_{\rm rel}^{-1}$ is produced by relaxation and
$Q_{\rm anh}^{-1}$ by anharmonicity, the total
$Q^{-1}$ will simply be $Q_{\rm rel}^{-1} + Q_{\rm anh}^{-1}$.
It should also be noted that the right-hand side of (1) only describes velocity changes
associated with frequency dependent damping, while ``static'' velocity changes
$(v_\infty(T)-v_0)/v_0$ might result from $T$-dependent structural effects.

The paper is organized as follows.
In Section II, the experimental aspects are described, including the
spectrometer and the characterization of the sample.
In Section III, the new Brillouin scattering data obtained on
a well densified sample of $d$-SiO$_2$ are presented.
These are then analyzed in Section IV.
A discussion concludes the paper in Section V.

\section{Experimental aspects}

The experiments reported here
are particularly demanding both in terms of sample quality and
spectrometer resolution.
The Brillouin frequency shift $\nu _ {\rm B}$ produced by
light scattering from an acoustic wave of velocity $v$ in an isotropic
medium is given by
$$ \nu _ {\rm B} \; = \;
\frac{2 n v}{\lambda_{\rm L}} \; {\rm sin}\; (\theta /2)\; \;,\eqno{(3)}$$
where $n$ is the refractive index of the sample, $\lambda_{\rm L}$ the
vacuum wavelength of the exciting laser, and $\theta$ the scattering
angle.
The scattered light must be collected over a finite aperture $\Delta \theta$.
Hence, if one is interested in small linewidths -- which is eminently
the case here --
the broadening produced by $\Delta \theta$, proportional to
cos$(\theta / 2) \; \Delta \theta$, must be minimized
by working with the smallest acceptable $\Delta \theta$ and near
backscattering, {\em i.e.} for $\theta \simeq 180^\circ$.
In isotropic media the transverse acoustic modes are not
active in backscattering.
This is why the measurements are restricted here to the LA-mode.

Densification has an enormous effect on $v$ which at room $T$ changes from
$\sim 5950$ m/s for $v$-SiO$_2$ to $\sim 7050$ m/s for $d$-SiO$_2$
with $\rho = 2.63$ g/cm$^3$.
Thus, near backscattering and at
$\lambda_{\rm L} = 5145$ \AA, $\nu _ {\rm B}$ changes
from $\simeq 34$ GHz in $v$-SiO$_2$ to $\simeq 42$ GHz in $d$-SiO$_2$.
On the other hand, densification also reduces considerably the sound
attenuation and hence the half-width of the Brillouin line.
In $v$-SiO$_2$ at room $T$ a half-width
$\Delta \nu _{\rm B} \simeq 75$ MHz is observed \cite{Vac81},
whereas it falls below 20 MHz in $d$-SiO$_2$.
Note that $\Delta \nu  _ {\rm B}$ would equal $\Gamma / 2 \pi$  -- the
quantity of main interest here -- in absence of parasitic
sources of broadening.
In fact, these narrow linewidths can only be
measured on samples that are very homogeneous in their density.
It is easily estimated from the above figures that an inhomogeneity
within the scattering volume  of only 1\% in density would lead
to a broadening of the Brillouin signal by about 400 MHz.
This value would be totally unacceptable when the instrinsic width
of interest is only a few MHz.
Furthermore, such small linewidths are usually not measurable in
commercial Brillouin instruments.
The Brillouin spectrometer used in these measurements is briefly described 
in Subsection II A.

Permanently densified silica, $d$-SiO$_2$, is obtained by submitting
$v$-SiO$_2$ to high pressure at elevated $T$ \cite{Bri53}.
Our series of samples was prepared from short cylinders of silica
that were
submitted to $\simeq 8$ GPa of quasi-hydrostatic pressure at
$T \simeq 700$ $^\circ$C.
Different densities were achieved by varying the duration of the treatment.
We used samples with initial $\rho$ values of 2.31, 2.46, and
2.63 g/cm$^3$, besides $v$-SiO$_2$ with $\rho = 2.20$ g/cm$^3$.
A typical raw sample had a diameter of 4 mm and a length of 8 mm.
Although the samples are often cracked, those at the highest densities 
can be clear and transparent.
Their surfaces are smooth but not flat.
Thus, to perform Brillouin scattering at room $T$, the samples can be
placed in an index matching fluid.
This is not feasible at either cryogenic or elevated temperatures.
Hence, a selected sample of the highest density had to be cut and
polished with flat faces.
It was then carefully characterized to identify a suitable
region to perform these experiments.
This is described in Subsection II B.

\subsection{High resolution spectrometer}

Three main criteria are to be satisfied by a spectrometer suitable for
these Brillouin measurements: contrast, resolution, and accuracy.
A very high contrast between the Stokes shifted frequency $\nu_{\rm S} =
\nu_{\rm L}-\nu_{\rm B}$ and the exciting laser frequency $\nu_{\rm L}$
can be achieved by multiple passes through a planar Fabry Perot (PFP)
interferometer \cite{San71}.
However, the resolution of such a PFP is insufficient to measure
$\Delta \nu_{\rm B}$.
It is thus placed in series with a spherical Fabry Perot (SFP) which is the
scanned instrument.
The high accuracy on $\nu_{\rm B}$ is achieved by producing a frequency
adjustable
reference signal  $\nu_{\rm M} \simeq \nu_{\rm S}$ with an electro-optic
modulator working near $\nu_{\rm B}$.
In practice, this reference is also used to dynamically adjust the PFP
which works as a filter only transmiting the Stokes Brillouin line.

\begin{figure}
\includegraphics[width=8.5cm]{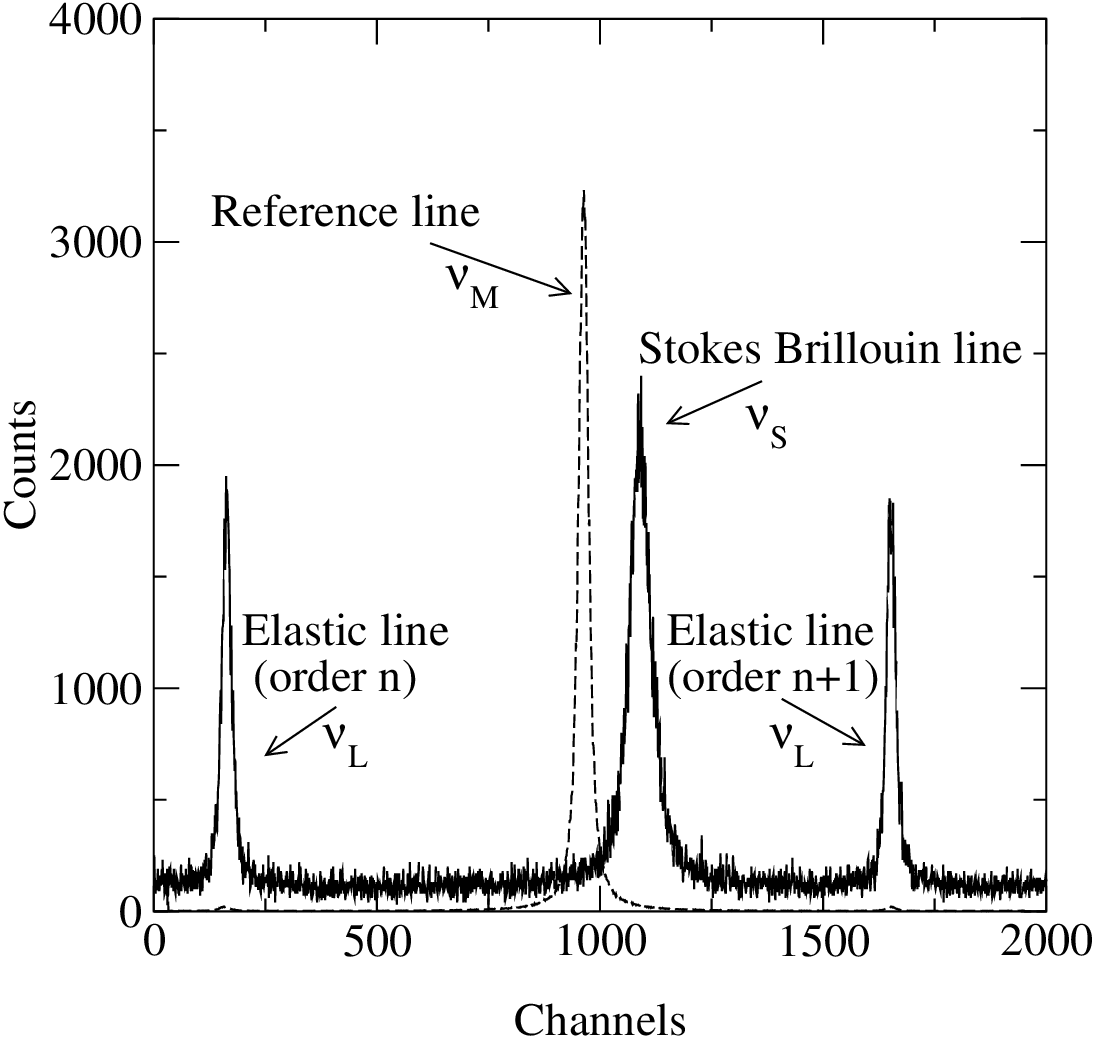}
\caption{Signal obtained in scanning a full order of the spherical Fabry-Perot.
One recognizes two elastic peaks marked $\nu_{\rm L}$, the Stokes shifted peak $\nu_{\rm S}$,
and the modulation signal $\nu_{\rm M}$.
The latter is stopped during the acquisition of the Stokes signal.}
\end{figure}
We used a modern version of the instrument described in \cite{Vac80b}.
The PFP has a spacing of 2 mm and it is used in four passes.
The SFP has a free spectral range of 1.5 GHz and a finesse of $\sim 40$.
A typical spectrum obtained in scanning just the SFP is illustrated in
Fig. 1.
One recognizes two orders of the elastic line at $\nu_{\rm L}$, the
Stokes Brillouin signal at $\nu_{\rm S}$,
and the modulation signal at $\nu_{\rm M}$.
All these lines are at a rather high interference order.
A very high accuracy on $\nu_{\rm S}$ is obtained as only the fractional
order separating it from $\nu_{\rm M}$ needs to be determined,
while $\nu_{\rm M}$ itself is known to the extremely high accuracy of
the frequency generator.
This provides for a determination of $\nu_{\rm B}$ to better than 2 MHz
absolute accuracy provided the Brillouin signal is appropriately
intense and symmetric.
The instrumental half-width of this SFP is $\simeq 20$ MHz.
Using photocounting and numerical deconvolution the half-width of
a sufficiently intense
damped harmonic oscillator (DHO) Brillouin line can then be determined
with $\pm 2$ to 3 MHz precision.

\subsection{Sample characterization}

The mean densities of the various samples were carefully determined by the
flotation method.
Their refractive index at the sodium D lines ($\lambda = 5893$ \AA)
was measured with an Abbe refractometer.
This gave results in good agreement with a previous determination
\cite{Arn88}.
The corresponding indices for $\lambda_{\rm L} = 5145$
\AA\ were extrapolated.
Their values vary linearly with $\rho$ and are given by
$$ n(5145 \; {\rm \AA}) \; = \; 1.022 \; + \; 0.200 \; \rho \; \;
{\rm (g/cm^3)} \; \; , \eqno{(4)}$$
to $10^{-3}$ accuracy.

\begin{figure}
\includegraphics[width=8.5cm]{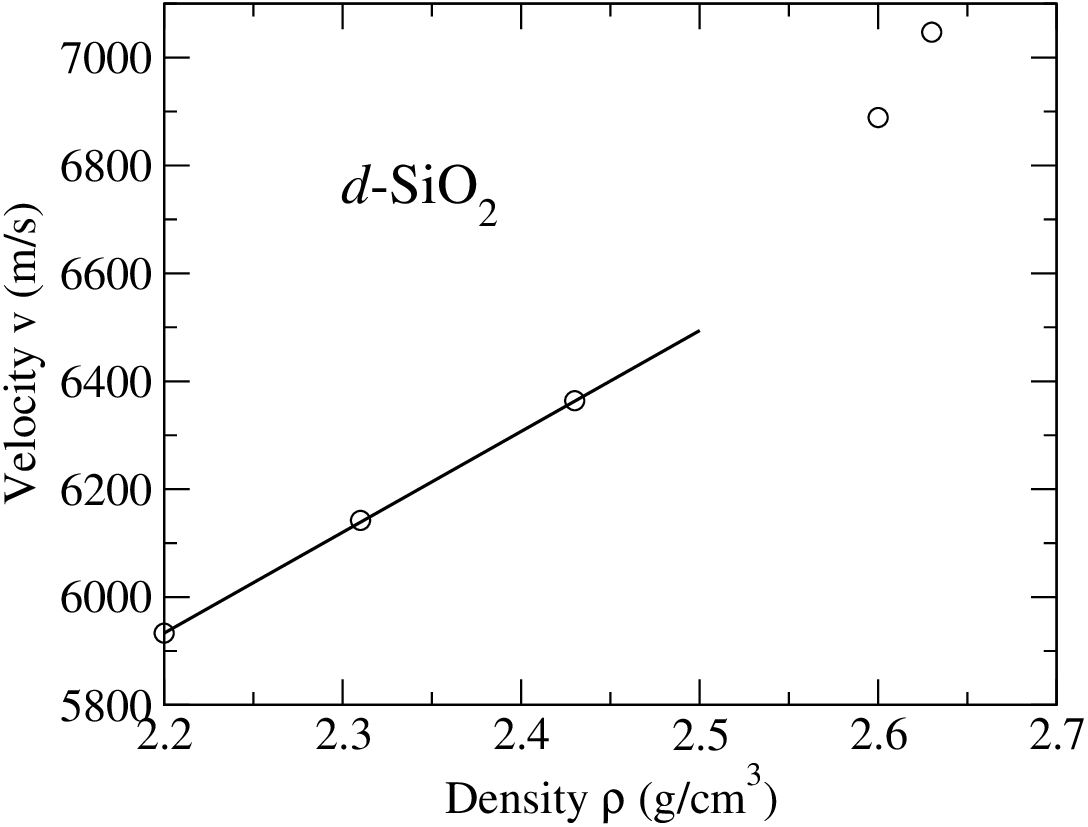}
\caption{Hypersonic velocity measured in $d$-SiO$_2$ in function of the sample density.
The line is a guide to the eye showing the initial slope.} 
\end{figure}
Figure 2 shows the hypersonic velocities measured at room temperature
on the four original densities, plus one at an additional density
of 2.60 g/cm$^3$ (see below).
The samples of intermediate densities (2.31 and 2.46) were much cracked
internally, which produced a spread of velocities that translated
into broader and distorted Brillouin lines.
This did not prevent a sufficiently accurate determination of a mean
velocity, with a $\pm 10$ m/s error bar which is still small on the
scale of Fig. 2.
The important point is that $v$ is a strong function of $\rho$.
One finds ${\rm d}v/{\rm d}\rho = 1.87 \times 10^3$ in m/s per g/cm$^3$,
at small densifications.
As seen in Fig. 2, this slope steepens
towards higher densities.

Brillouin scattering is a rather local probe.
In the backward geometry used here, different small areas of a sample
cross-section can be probed.
The measurement of $\nu_{\rm B}$ is then
a test for the sample homogeneity.
This revealed that one of the samples of highest density was
rather homogeneous, with $\nu_{\rm B}$ ranging from 42.32 to
42.35 GHz across its section.
This piece was then selected for cutting and polishing, to obtain flat
faces in order to be able to perform the Brillouin measurements in
function of $T$.
Unfortunately, this mechanical process introduced an appreciable distribution
of frequencies, to such an extent that the Brillouin lines became
quite asymmetric.
This sample was then used for Brillouin scattering
with x-rays in \cite{Rat99}, and subsequently
in \cite{For02} and \cite{Ruf03}.
It should be remarked that these inhomogeneities do not appreciably
affect the x-ray Brillouin scattering results.
The latter experiments are indeed performed with four orders of magnitude
less frequency resolution than here, the error bars on
$\Gamma / 2 \pi$ being then at best $\pm 30$ GHz \cite{Ruf03}.
During the x-ray measurements the sample was maintained at
an elevated temperature for a long time in order to increase the signal
by the thermal population of the acoustic mode.
We used $T=575$ K \cite{Rat99}, a temperature which is too low to appreciably
relax the densification over the long duration of the x-ray experiment.
Checking the optical Brillouin spectrum after the first x-ray experiment we
discovered that the sample had become locally much more homogeneous,
presumably owing to the thermal treatment.
Nice symmetric and narrow Brillouin lines were then observed.
At room $T$, one end of the sample had a velocity indicating $\rho
\simeq 2.60$
g/cm$^3$, while near the opposite end we found $\rho \simeq 2.57$.
The corresponding difference in $\nu_{\rm B}$ is of the order of 1 GHz.
The former region had a linewidth $\Delta \nu_{\rm B} = 25$ MHz, while the
latter had $\Delta \nu_{\rm B} = 40$ MHz.
These two regions were separated by a less homogeneous one,
with asymmetric Brillouin lines and a higher width.
For the measurements presented in Section III, the region of density
2.60 and with the narrower line was selected.
It should be remembered that part of the observed broadening can easily
arise from small remaining inhomogeneities.

\section{Brillouin scattering results}

Figure 3 presents the Brillouin results obtained on the sample region
with $\rho = 2.60$ as explained in Subsection II B.
For the low-$T$ measurements, the
sample was placed in a He-flow optical cryostat.
The temperature of the gas flow is stabilized to much better than 1 K,
and the sample temperature is read with a silicon diode.
For the high-$T$ measurements, the sample was mounted in an optical oven
stabilized to $\pm 1$ K.
The value of $\theta$, the scattering angle internal to the sample,
was taken at $178.1^\circ$ for these measurements in order to minimize
the collection angle broadening.
The measured spectra were fitted to a DHO response function of frequency
$\nu_{\rm B}$ and half-width $\Delta \nu_{\rm B}$, convoluted
with the experimental instrumental function.

\begin{figure}
\includegraphics[width=8.5cm]{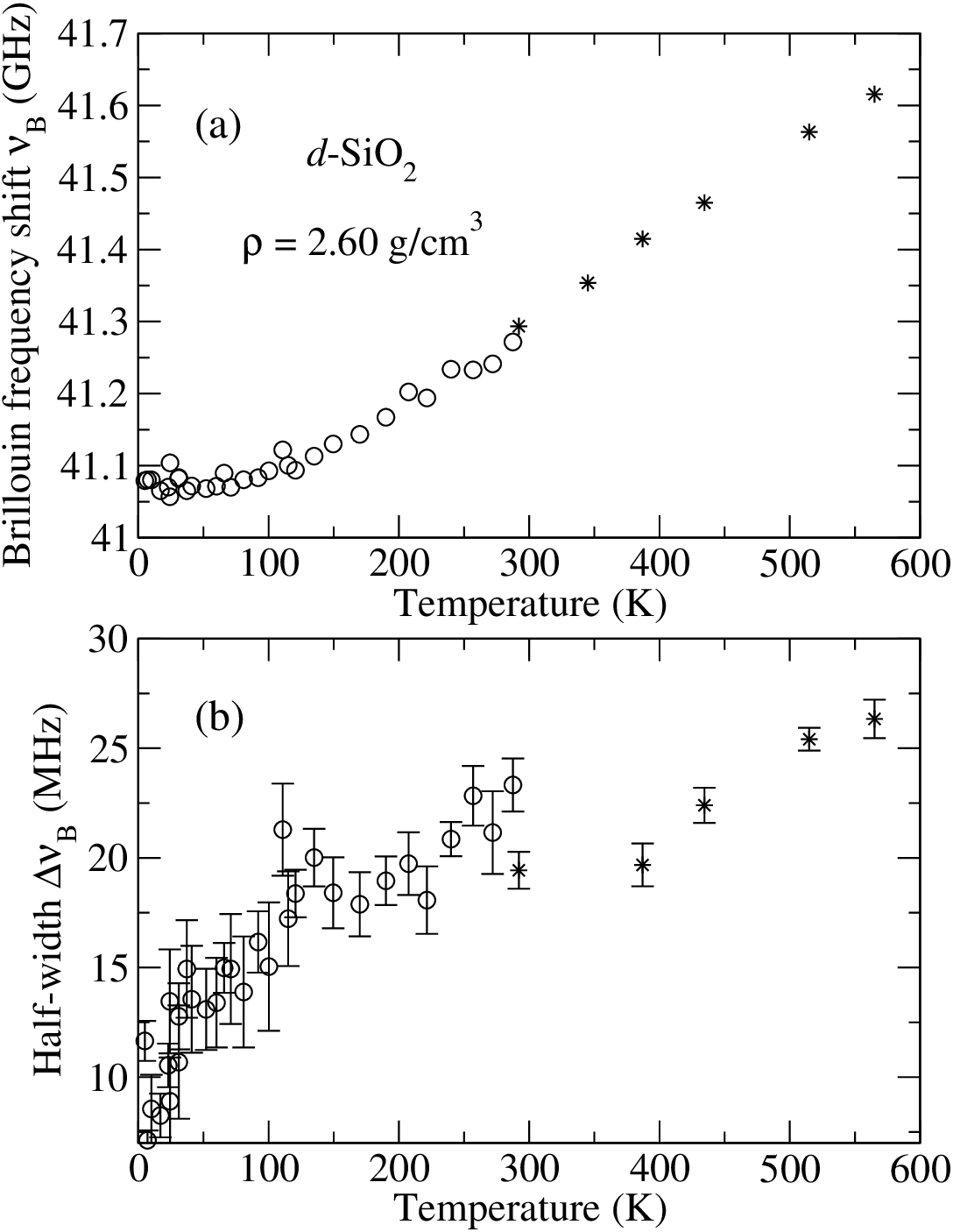}
\caption{The measured Brillouin frequency shift $\nu_{\rm B}$ (a), and the half-width $\Delta \nu_{\rm B}$ (b),
in function of the temperature $T$ for the sample of density $\rho = 2.60$ g/cm$^3$.
Different symbols are used for data taken in the cryostat and in the oven.
Although the raw data in (b) might suggest a dip around room temperature, it is more likely that this
effect is related to sample inhomogeneities, as explained in the text.}
\end{figure}
The values $\nu_{\rm B}$ are shown in Fig. 3a.
The error bars on these values are of the order of the size of the points.
The values $\Delta \nu_{\rm B}$ are shown in Fig. 3b with error bars
that are given by the Marquardt-Levenberg least-square fitting algorithm.
Care was taken to keep the laser beam on the same sample spot during the several
days needed for measurements.
It was however not possible to recover exactly
the same position on the sample after
its transfer from the cryostat to the oven.
If residual inhomogeneous broadening is present -- as we suspect --
its amount is likely to vary from place to place.
This could explain the difference -- beyond the size of the error bars --
between the level of $\Delta \nu_{\rm B}$ observed in the cryostat
and that in the oven.

\begin{figure}
\includegraphics[width=8.5cm]{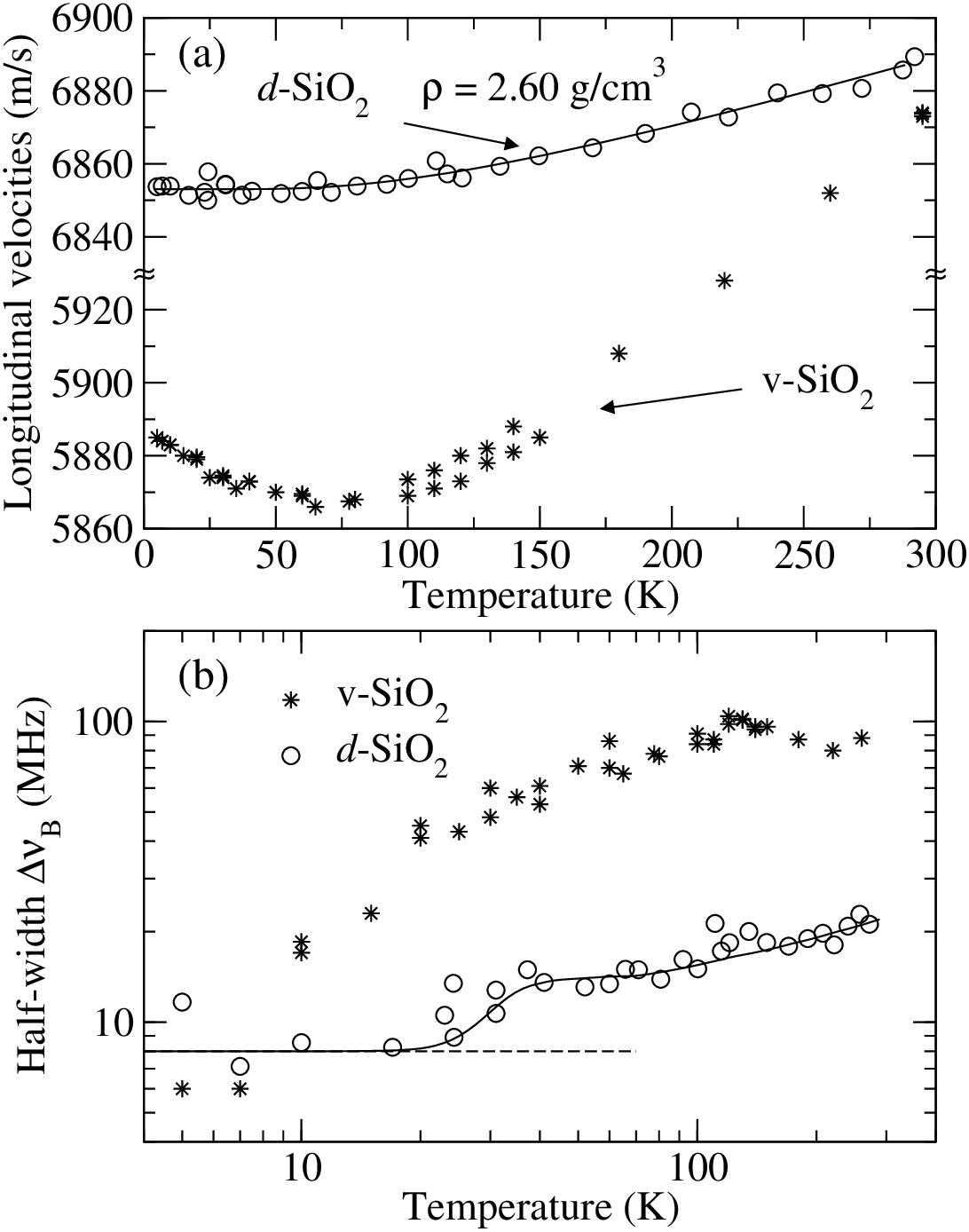}
\caption{Comparison of $d$-SiO$_2$ and $v$-SiO$_2$ showing the velocities (a) and the linewidths (b) below room temperature.
The $v$-SiO$_2$ data was acquired long ago [17] with a similar spectrometer but at a laser wavelength of 4880 \AA $\;$
and without electro-optic modulated reference.
The solid lines result from calculations explained in the text.
The dashed line in (b) indicates the estimated inhomogeneous background of the $d$-SiO$_2$ sample.}
\end{figure}
It is now of interest to briefly  compare the results obtained on
$d$-SiO$_2$ to these in $v$-SiO$_2$.
Fig. 4 illustrates data obtained long ago on $v$-SiO$_2$ \cite{Sil81}.
The velocities below room-$T$ are compared in Fig. 4a.
The values of $v$ for $d$-SiO$_2$ are derived from the data shown in Fig. 3a 
using Eq. (3).
As $T$ is raised from liq. He temperatures, one clearly observes a dip
of $v$ in $v$-SiO$_2$.
Within the accuracy of our experiment, such a dip is totally
absent in $d$-SiO$_2$.
It is well known that the dip is produced by the $T$-dependence of
thermally activated  relaxational motions.
This will  further be discussed in paper (II).
Its absence in $d$-SiO$_2$ demonstrates that thermally activated
relaxation becomes negligible after densification.
A strong decrease of these relaxations compared to $v$-SiO$_2$ has
already been reported for the less densified sample of $d$-SiO$_2$
that was examined with both Brillouin scattering \cite{Doe94}
and vibrating reed \cite{Wei96} measurements.
In that sample there remained a clear dip in $v$, but the relaxation
peak observed with vibrating reed had decreased by a factor of about six
compared to $v$-SiO$_2$.
We conclude that in our case, the dip in $v$ being absent, the
relaxation contribution to the damping must be totally negligible.

Fig. 4b shows the corresponding half-widths. A logarithmic scale is used here to
emphasize the low-$T$ region and also because the damping in
$v$-SiO$_2$ is quite a bit larger than in $d$-SiO$_2$.
As discussed in paper (II), the $v$-SiO$_2$ values result from a
combination of relaxation and anharmonicity.
Below 10 K, they fall however below the $\Delta \nu_{\rm B}$ values observed
in $d$-SiO$_2$, which would be surprising were it not for the inhomogeneity
of the $d$-SiO$_2$ sample dicussed above.
Hence, we attribute the level of 8 MHz, illustrated by the dashed line,
to the residual inhomogeneous background.
The corresponding level for the data measured in the oven is estimated to be
about half of that, {\em i.e.} $\sim 4$ MHz.

\section{Network viscosity}

The data shown in Fig. 3b, corrected for the estimated inhomogeneous
background,
is used to calculate the internal friction $Q^{-1} = 2 \Gamma / \Omega$
shown in Fig. 5.
\begin{figure}
\includegraphics[width=8.5cm]{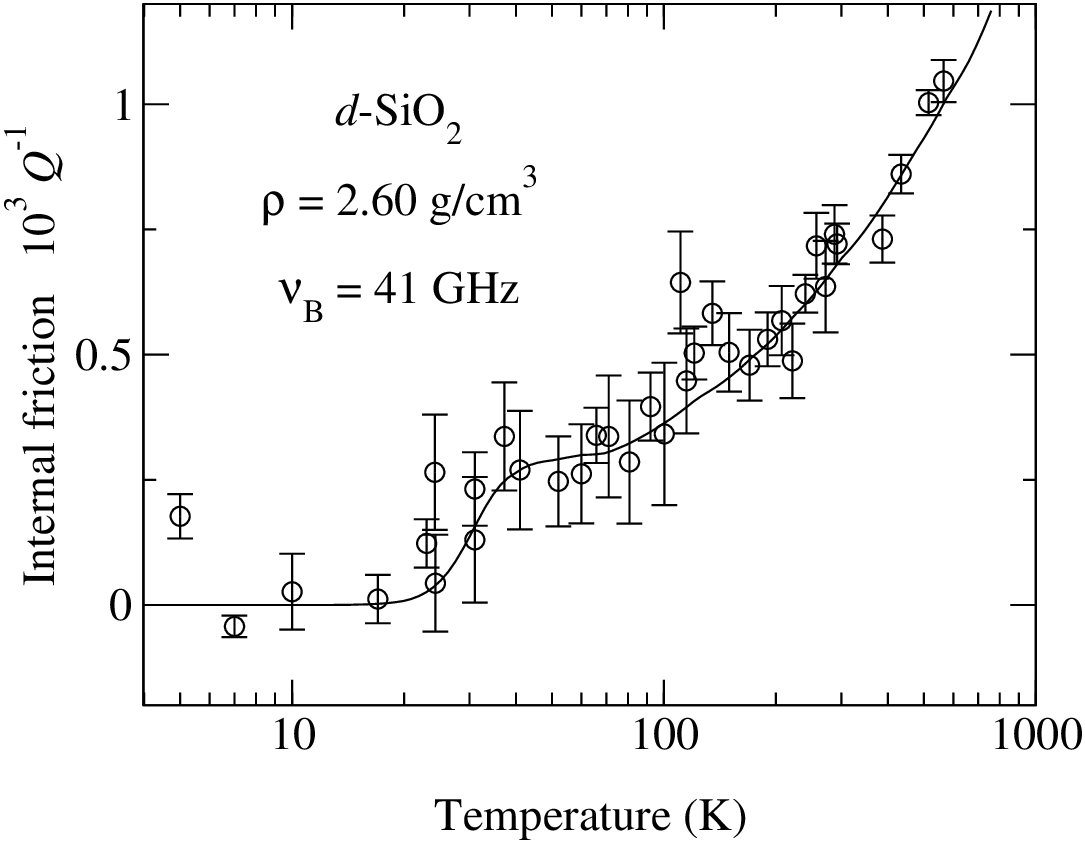}
\caption{The internal friction of the $d$-SiO$_2$ sample. The calculated line is explained in the text.}
\end{figure}
As explained above, this result should be interpreted in terms of the
$Q^{-1}_{\rm anh}$ produced by anharmonicity.
The theory of lattice viscosity was developed for crystals,
in which case the calculation of correlation functions \cite{Vau67b}
or the sums over modes \cite{Mar71,Woo61}
could be carried out quite far.
For glasses, only the fracton model has been pursued to the point of
making analytic predictions for the acoustic linewidth \cite{Ale86,Jag89}.
However, the development described in \cite{Mar71}, in particular
the one leading to Eq. (131) of that review, is essentially based on the
standard equations of viscoelasticity supplemented by a simplifying
assumption regarding the details of the collision term in the Boltzmann
equation.
A fully parallel reasoning can be made for glasses.
Hence, for $\Omega \tau_{\rm th} \ll 1$, similarly to the phonon
viscosity term of \cite{Mar69}, we posit that network viscosity in glasses
leads to the internal friction
$$Q^{-1}_{\rm anh} \; = \; A \Omega \tau_{\rm th} \; \; ,\eqno{(5a)}$$
with
$$ A \; = \; \gamma ^2 \frac{C_{\rm v} T v} {2 \rho v_{\rm D}^3} \;
 \; . \eqno{(5b)}$$
Here $C_{\rm v}$ is the specific heat per unit volume, $v$ is the
velocity of the sound wave, $v_{\rm D}$ is the Debye velocity, and
$\gamma ^2$ is a mean squared average Gr\"uneisen parameter
for the thermally excited modes \cite{Mar71,Fab99}.
We neglect the thermal conduction term in Eq. (131) of \cite{Mar71}.
Experience shows that the latter is already considerably smaller than
the phonon-viscosity one in crystals \cite{Bon83}.
This should be even more so in glasses owing to the much reduced thermal
conduction \cite{Fab99}.
The strongest assumption in (5) is presumably the use of a single
relaxation time $\tau_{\rm th} (T)$.
A second assumption is to take $\gamma ^2$ constant.

As the temperature is lowered, one expects that  $\tau_{\rm th}$
increases, so that the condition $\Omega \tau_{\rm th} \ll 1$ 
eventually becomes violated.
We use the Ansatz that $Q^{-1}_{\rm anh}$ is then given by (2a), where
$\tau = \tau_{\rm th}$ and $A$ is given by (5b),
$$ Q^{-1}_{\rm anh}= A \Omega \tau_{\rm th} / (1+ \Omega ^2 \tau_{\rm th} ^2 ) \; , \eqno{(6)}$$
As pointed out in \cite{Vau67b}, this amounts to introducing an
exponential decay in the time-dependent relaxation function.
This is preferable to the expression (4.15) proposed in \cite{Woo61},
as the latter does not have a finite Kramers-Kronig transform.

It is now of interest to estimate $\tau_{\rm th}$ from the data shown in Fig. 5.
This is done using (6), in which all quantities are quite well known except
for the value of $\gamma ^2$ that enters $A$ according to $(5b)$.
The other non-trivial quantities in $A$ are: 1) the specific heat $C_{\rm v}(T)$ for which we use
the data from \cite{Ina99}, complemented at high-$T$ with the specific heat
of $v$-SiO$_2$ appropriately scaled to account for the higher density of $d$-SiO$_2$; 
2) the Debye velocity for which we use $v_{\rm D}^3 = 0.322 \; v^3$, a numerical
coefficient quite universal for glasses and taken from $v$-SiO$_2$.
With the above, one calculates $A/\gamma ^2$, which is a quantity that increases rapidly with $T$
at low $T$.
One divides then $Q^{-1}$ by that quantity, to obtain a result which according to (6)
is equal to $\gamma ^2 \Omega \tau_{\rm th} / (1+ \Omega ^2 \tau_{\rm th} ^2 )$.
One notes that in function of $T$ the latter is maximum
at $\Omega \tau_{\rm th} = 1$, and that the maximum value is simply $\gamma ^2 /2$. 
Unfortunately the noise in $Q^{-1}/(A/\gamma ^2)$  becomes much too high at low temperatures
owing to the smallness of $A/\gamma ^2$, so that a clear maximum cannot
be identified.
One notes however that the maximum in $Q^{-1}_{\rm anh} /A$ is not a maximum in
$Q^{-1}$, as $A$ is a rapidly growing function of $T$.
At $\Omega \tau_{\rm th} = 1$, there is rather a step-like increase in $Q^{-1}(T)$.
Such a feature is clearly recognizable between 25 and 35 K in the semilog
presentation of Fig. 5.
This suggests that $\Omega \tau_{\rm th}$ reaches 1 around 30 K.
An extrapolation of $Q^{-1}_{\rm anh} /A$ from higher $T$ indicates then that
$\gamma ^2 \approx 8$ is an appropriate value, with an uncertainty smaller than 20 \%.
The resulting value $\gamma \simeq 2.8$ is reasonable, especially owing to 
the observations made in \cite{Fab97,Fab99}.
To complete the task,
one can then solve for $\Omega \tau_{\rm th}$ the quadratic equation (6).
For each data point, the lowest root must be kept for
$\Omega \tau_{\rm th} \leq 1$ and
the highest one for $\Omega \tau_{\rm th} \geq 1$.
The noise in our data produces some complex roots for $T < 40$ K. These have been rejected.
Dividing the roots by $\Omega$, the values of $\tau_{\rm th}$ shown in Fig. 6
are then obtained.
The uncertainty in $\gamma ^2$ produces a similar uncertainty in the absolute scale of $\tau_{\rm th}$.
This is of no consequence for the following discussion.
\begin{figure}
\includegraphics[width=8.5cm]{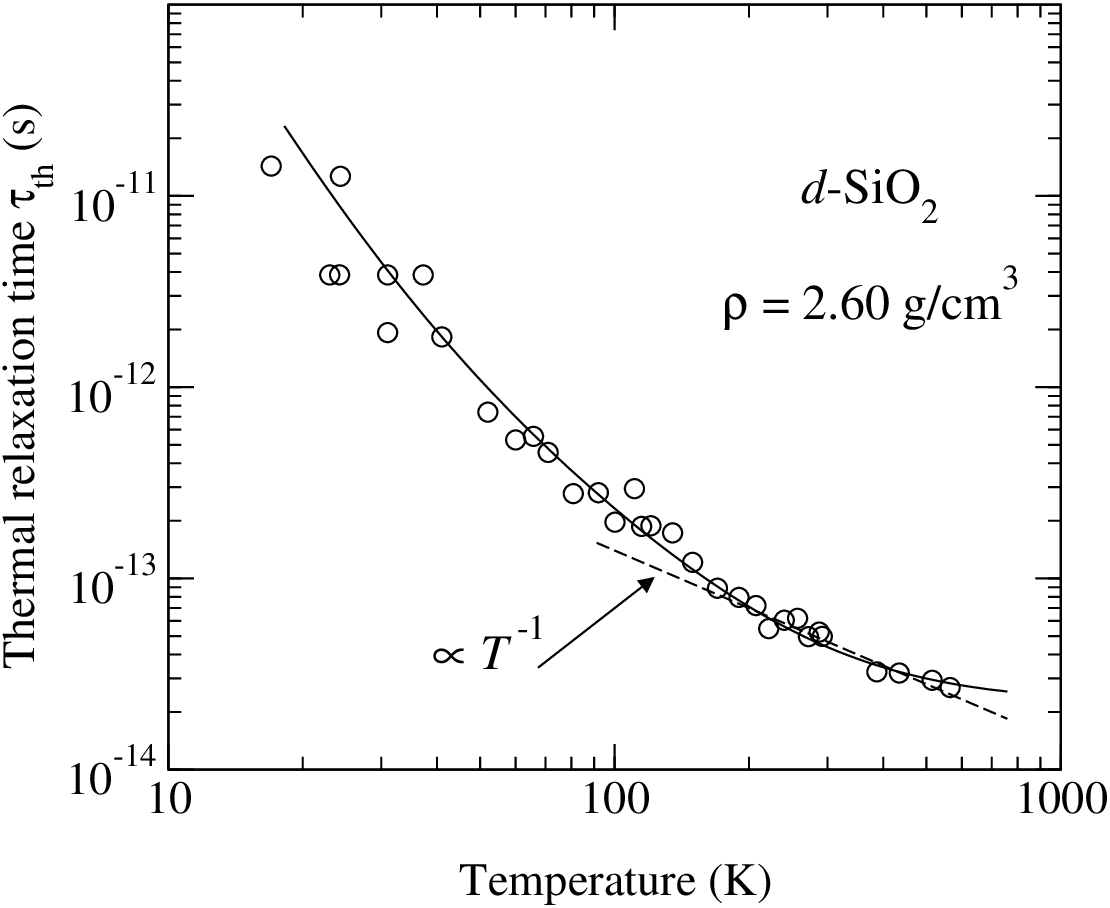}
\caption{The thermal relaxation time obtained from the data in Fig. 5 using eq. (5) and (6) as explained in the text.
The dashed line shows the $1/T$ behavior at high $T$.
The solid line is a guide to the eye explained in the text.}
\end{figure}

We observe that $\tau_{\rm th}$ increases rapidly on cooling.
It varies approximately in $1/T$ from $\sim 150$ to 600 K, and progressively faster
below $\sim 100$ K.
This is illustrated by the dashed line of slope -1 in Fig. 6.
To produce the guides to the eye used in Figs. 4 to 6, we first adjusted
$\tau_{\rm th} (T)$ to a sum of inverse powers of $T$, obtaining the
solid line shown in Fig. 6.
We also adjusted $v(T)$ to
$v_0 [ 1 - a / (e^{\Theta / T} - 1)]$ which is derived from
Wachtman's equation as discussed in \cite{And66}.
This gives $v_0 = 6853$ m/s, $a = - 1.052 \times 10^{-2}$,
and $\Theta = 328$ K, corresponding to
the line shown in Fig. 4a which provides an excellent
fit of $v$ up to the highest measured data point at $T$ near 600 K.
Having $v(T)$, we can calculate $\Omega (T)$, and then  $Q^{-1}_{\rm anh} (T)$
and $\Delta \nu_{\rm B} (T)$ as shown in Figs. 5 and 4b.
The sharpness of the step near 30 K strongly depends on the assumed
$T$-dependence of $\tau_{\rm th} (T)$ in that region.
Its exact position also depends on the value selected for  $\gamma ^2$.
At this stage, the limited quality of the data would not justify
attempting any further improvement on these adjustments.
The main difficulty obviously arises from the inhomogeneous background
contribution to $\Delta \nu_{\rm B}$.

We can now use Eq. (2) to calculate the velocity change
$(\delta v / v)_{\rm anh}$ associated with $Q^{-1}_{\rm anh}$.
As indicated in the Introduction, the result of this calculation
is then referred to the unrelaxed velocity $v_{\infty}(T)$ whose $T$-dependence can have a structural origin.
Thus one has
$$\frac{v-v_{\infty}(T)}{v_{\infty}(T)} \; = \; - \frac{A/2}
{1+\Omega^2 \tau_{\rm th}^2} \; \;  , \eqno{(7)}$$
where $v$ is the Brillouin value measured at frequency $\Omega$.
All velocity changes being relatively small,  $v_{\infty}(T)$ can be replaced
by  $v_{\infty}(0)=v_0$ in the denominator on the left-hand side.
Further, $A$ being anyway very small when $\Omega \tau_{\rm th} \geq 1$,
$\Omega \tau_{\rm th}$ can be neglected in the denominator on the
right-hand side of (7), so that finally
$$\frac{v_{\infty}(T)-v_0}{v_0} \; \simeq \; \frac{v-v_0}{v_0}+
\frac{A}{2} \; \;  . \eqno{(8)}$$
We remark that the size of the ``correction'' term $A/2$ is directly
proportional to the value selected for $\gamma ^2$.
\begin{figure}
\includegraphics[width=8.5cm]{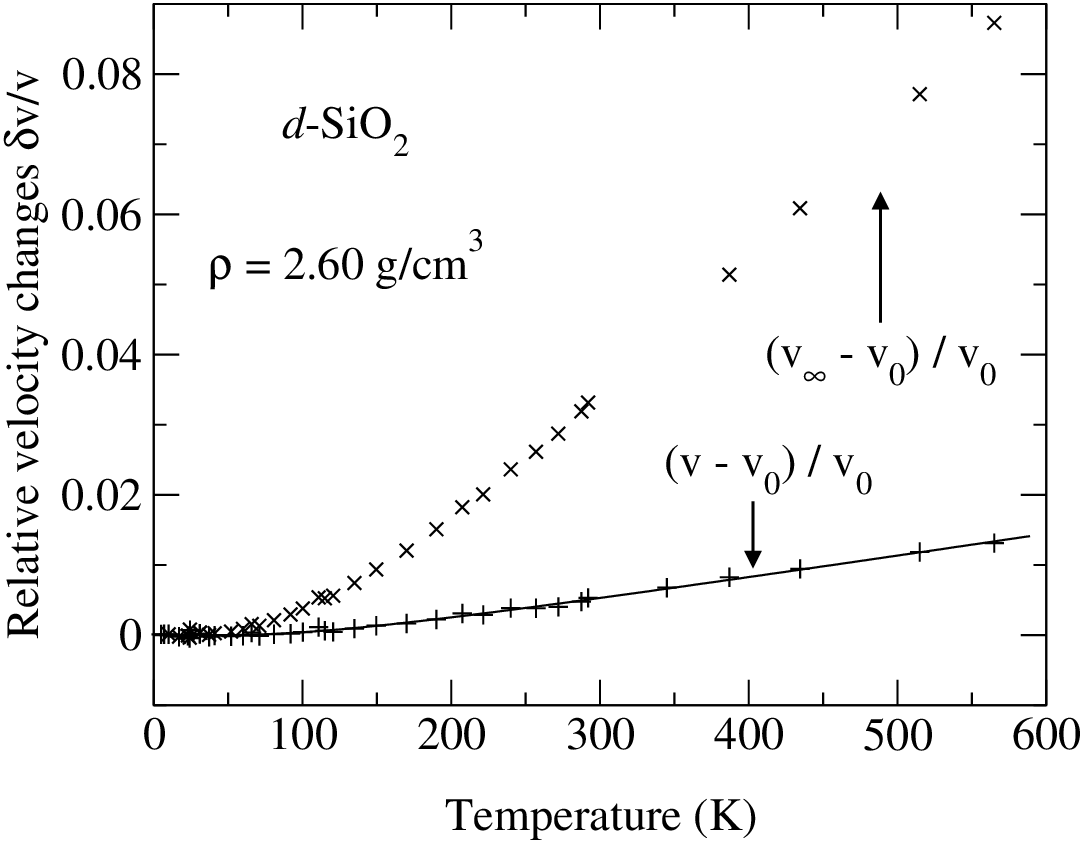}
\caption{The relaxed $(v-v_{\rm 0})/v_{\rm 0}$ and unrelaxed $(v_\infty-v_{\rm 0})/v_{\rm 0}$ velocity changes.
The line is the adjustment with Wachtman's equation \cite{And66}.}
\end{figure}
Figure 7 shows $(v-v_0)/v_0$ and $(v_{\infty}-v_0)/v_0$.
Both exhibit an anomalous $T$-dependence, increasing as $T$ increases,
but this anomaly becomes quite a bit stronger for the
unrelaxed velocity $(v_{\infty}-v_0)/v_0$.

\section{Discussion}

The above results suggest that in $d$-SiO$_2$, contrary to all other
investigated glasses, the damping of sound observed in Brillouin scattering
is purely produced by the network viscosity, {\em i.e.} by the
anharmonic interaction of collective modes rather than by the
relaxation of local ``defects''.
Two features deserve additional discussion:
1) the temperature dependence of the relaxation time of the dominant thermal modes, $\tau_{\rm th}(T)$;
2) that of the unrelaxed sound velocity $v_{\infty}(T)$.

The particular behavior of $\tau_{\rm th}(T)$, which decays $\sim 1/T$ above $\sim 150$ K,
probably reflects the fact that in a glass, as
opposed to crystals, the thermal modes are mostly diffusive
or non-propagating.
By comparison, in crystal quartz for example there is a large range of
$T$ where $\tau_{\rm th}$ decreases with $1/T^2$ \cite{Boe60}.
To understand the nature of this difference, we consider the only available calculation
of anharmonic lifetimes in disordered systems.
It is based on the fracton model \cite{Ale86}.
This model accounts for the non-propagating nature of the modes.
As pointed out by the authors of \cite{Jag89}, fractons have been used to
``obtain explicit results'', but ``the approach would apply to any
localized vibrational states''.
We believe this remark is in particular true for what concerns the
functional dependence in $T$ of the relaxation times.
Indeed, the dependence found in \cite{Ale86} does not involve
any of the various fractal dimensions, but just the crossover energy
$\hbar \omega _{\rm co}$ above which the modes become non-propagating.
The $T$-dependence results from the thermal population of the modes and
their spatial overlap.
It is presumably of quite general applicability to such situations,
independently from the fractal model.
The crossover energy is well known for $d$-SiO$_2$ as it has been
carefully determined by x-ray inelastic scattering
measurements \cite{Rat99,For02,Ruf03}.
It is located at about 9 meV, or correspondingly at
$T_{\rm co} = \hbar \omega _{\rm co} / k_{\rm B} \simeq 110$ K.
The prediction of \cite{Ale86} is that all relaxation times vary like
$\tau \propto 1/T$ for $T > T_{\rm co}$, whereas they diverge with
$\tau \propto {\rm exp}(T_{\rm co}/T) $ at lower temperatures $T < T_{\rm co}$.
We note that,
qualitatively, this is in perfect agreement with our observations.

Further, the thermal conductivity $\kappa$ in the Debye model
can be calculated from
$\tau_{\rm th}$ using the standard kinetic equation
$\kappa = C_{\rm v} v_{\rm D}^2 \tau_{\rm th} /3$ \cite{Kit76}.
Doing this, we find a calculated $\kappa$ which is considerably larger than the
measured one \cite{DMZ94}. Its $T$-dependence also
disagrees with the observations, as the calculation gives a $\kappa (T)$ that decreases with
increasing $T$ at high $T$, like in a crystal,
contrary to the observed increase \cite{DMZ94} beyond the typical thermal
conductivity plateau \cite{Zel71} common to glasses.
The reason for this is of course that the modes contributing
to $C_{\rm v}$ are at best diffusive so that the kinetic equation
is not applicable.
This provides a second and independent indication that the thermal modes are essentially non-propagating. 

Regarding the sound velocity, we first remark that the network viscosity has
the usual effect of decreasing the velocity from $v_\infty$ to $v$
as $T$ increases.
This is evident from a comparison of the two curves illustrated in Fig. 7.
What is abnormal is that the unrelaxed velocity $v_{\infty}(T)$
{\em increases} with $T$.
This anomaly is common to glasses that contain tetrahedral building blocks,
as shown experimentally for SiO$_2$, GeO$_2$, BeF$_2$, and zinc phosphate
in \cite{Kra68}.
An earlier proposal that this might be a manifestation of large
structural inhomogeneities special to refractory glasses \cite{Kul74}
does not seem supported by the fact that the effect is particularly
strong in BeF$_2$ which has a low glass transition temperature $T_{\rm g}$.
A more likely explanation might be that there is a specific structural feature
in SiO$_2$-like glasses which leads to a $T$-dependent reversible
structural change as shown in recent simulations \cite{Hua04}.
It was found in these that the angular position of the Si-O-Si plane around the Si-Si direction
has distinct equilibrium values separated by $90^\circ$ jumps.
The latter would produce in the glass a gradual transition similar in nature to the $\alpha - \beta$
transformation in cristobalite.
One might ask why there is no corresponding contribution to $Q^{-1}$
owing to such a transition-like mechanism.
A possible explanation it that the transition being distributed in $T$, most of the volume
is always very far from $\Omega \tau \sim 1$ so that the additional
damping might be much too weak to be observable.

Another question of interest is whether the unrelaxed velocity $v_{\infty}(T)$
could somehow become measurable in an appropriate experiment.
One could naively imagine that it would suffice to investigate very high
sound frequencies, as for example by going to sufficiently high momentum
exchange $Q$ in an x-ray Brillouin scattering experiment, to eliminate the
viscous relaxation term in (2b) by forcing $\Omega \tau_{\rm th} \gg 1$.
Consider however the value $\tau_{\rm th} \simeq 5 \times 10^{-14}$ s
which is found at room $T$ in Fig. 6.
To achieve $\Omega \tau_{\rm th} = 1$ would require
$\Omega / 2 \pi \simeq 3$ THz which is already above
$\omega _{\rm c} /2 \pi \simeq 2$ THz.
One runs then into a regime in which the sound waves are presumably
hybridizing \cite{Gur03} with the SiO$_4$ librational mode \cite{Heh00}, producing the
boson peak \cite{Cou03}.
Although such a measurement might give an indication about $v_{\infty}$,
it is certainly not a clean cut situation.
The other alternative would be to reduce the temperature so as
to increase the value of $\tau_{\rm th}$.
This also soon meets its limits as the size of the viscous relaxation
becomes negligible below $\sim 100$ K, as seen from Fig. 7.
We thus find that it would unfortunately be difficult to obtain a direct
measurement of the unrelaxed sound velocity.

To conclude, it seems that we have for the first time found a glass in which
the Brillouin sound damping appears to be entirely controlled by the
network viscosity. 
This already suggests that it is probably not appropriate to simply neglect
the anharmonic damping as proposed for silica in \cite{Wie00}.
This illustrates that the question is of importance, as it finally 
ties in with much more fundamental and debated issues, like the origin of the
thermal anomalies of glasses \cite{Phi81}.
We are fully aware that the noise in Fig. 2b is still far from satisfactory.
We could only progress on this issue if we had an appreciably more
homogeneous sample than the already excellent one used in this study.
If so, it would be relatively straightforward to increase the spectrometer
performance by using a spherical Fabry-Perot of larger spacing, such as
10 or 25 cm, the latter giving a factor five improvement over the present
resolution.

\end{document}